\documentclass{ws-procs9x6}            % default, citations in superscript
\usepackage{amsfonts}
\usepackage{amsmath}
\usepackage{amssymb}
\usepackage{natbib}
\usepackage{graphicx}

\def\beq{\begin{equation}}
\def\eeq{\end{equation}}
\def\bea{\begin{eqnarray}}
\def\eea{\end{eqnarray}}

\def\chic1{\chi_{c1}}

\def \MCDD {M_{\rm{CDD}}}

\begin{document}
\title{Nature of $X(3872)$ from the line shape}

\author{Xian-Wei Kang $^{1,2,*}$}

\address{$^1$ Key Laboratory of Beam Technology of Ministry of Education, College of Nuclear Science and Technology, Beijing Normal University, Beijing 100875, China\\
$^2$ Beijing Radiation Center, Beijing 100875, China\\
$^*$E-mail: 11112018023@bnu.edu.cn}

\author{J. A. Oller}

\address{Departamento de F\'\i sica,
Universidad de Murcia, E-30071 Murcia,
Spain\\
E-mail: oller@um.es}

\begin{abstract}
  Using a general parameterization of two-body scattering amplitude, we systematically analyze the corresponding data on $X(3872)$, more explicitly, the CDF data on inclusive $p\bar{p}$ scattering to  $J/\psi \pi^+\pi^-$, and the Belle and BaBar data on $B$ decays to $K\, J/\psi \pi^+\pi^-$ and $K D\bar{D}^{*0}$ around the $D^0\bar{D}^{*0}$ threshold.
We achieve a good reproduction of data and find
that the $X(3872)$ can be interpreted as a bound and/or virtual state, or even higher-order (double or triple)
  virtual sate pole.
The latter point was not realized previously in the literature.
%  The latter point has not been realized in other literatures.
As a result, the compositeness of the $X(3872)$ can vary largely from nearly 0 to 1.
More higher-precision data is needed to discriminate its pole structure and nature.
\end{abstract}

\keywords{CDD pole; two-body scattering; effective range expansion; line shape, compositeness; bound and virtual state.}

\bodymatter

\section{Introduction}\label{sec:Introduction}
A series of exotic hadron candidates, called XYZ particles because their properties
are hard to accommodate or cannot be accounted for by conventional potential models, have been observed.
A variety of theoretical interpretations have been proposed, e.g., kinematical effects, molecular states,
quark-gluon hybrids, etc.
We want to stress that, generally, different scenarios predict different constituents,
and in practice, it may involve several mechanisms.
In such circumstances the concept of compositeness is a valuable tool to obtain insight on the nature
of the resonance.
In most cases, those newly observed states stay close to threshold of hadron pairs \cite{Dai1,Dai2}
and the effective range expansion (ERE) may be applicable.
We have such a good example in Ref.~\cite{Zbpaper}, where the scattering length and effective range of $B\bar B^{(*)}$ are
calculated by reproducing the pole location of $Z_b$ states, and the molecular component of $B\bar B^{(*)}$ pairs
are predicted.
However, the ERE radius of convergence may be severely reduced in the presence of a Castillejo-Dalitz-Dyson (CDD)
pole \cite{CDD}.

\section{Formalism and data analysis}
Since the discovery of the $X(3872)$ by the Belle collaboration \cite{Belle},
a voluminous amount of papers devoted to its study have appeared.
Its quantum numbers have been well determined by now as $I^GJ^{PC}=0^+(1^{++})$ \cite{PDG}.
Within the ERE analysis, Ref.~\cite{Braaten} assigns the $X(3872)$ as a pure $D\bar D^{*}$ molecule,
while in Ref.~\cite{Hanhart} the virtual-state nature is proposed.

A nearby CDD pole around threshold could spoil the applicability of ERE to study the $X(3872)$ \cite{Hanhart2}.
In the presence of a CDD pole, the scattering amplitude can be written as \cite{Kang3872}
\begin{align}
\label{211116.3}
t(E)=&\left(\frac{\lambda}{E-M_{\rm CDD}}+\beta-ik(E)\right)^{-1}~,
\end{align}
where the CDD poles lies at $M_{\rm CDD}$, its residue is $\lambda$, $\beta$ is the subtraction constant
and $k(E)=\sqrt{2\mu (E+i\frac{\Gamma_*}{2})}$.
Throughout the paper, energy $E$, $\MCDD$ and pole position are defined with respect to the $D^0\bar D^{*0}$ threshold.
The width of $D^{*0}$ meson is $\Gamma_*\approx 65$ keV, and is taken into account in our formalism by
employing a complex mass for the $D^{*0}$.
This width may play an important role in a quantitative analysis, since the width of the $X(3872)$,
$\Gamma_X$, is also very small (smaller than 1.2 MeV as  determined by experiment \cite{PDG}).
Removing the extra zero due to the CDD pole, one ends with the function $d(E)$, which
accounts for the final-state interactions,
\begin{align}
\label{211116.7}
d(E)=&\frac{1}{1+\frac{E-\MCDD}{\lambda}(\beta-ik)}.
\end{align}
Equation~\eqref{211116.3} can be verified in the case
in which cross-channel effect \cite{Oller1999} can be disregarded, so that the partial-wave
amplitude only has a right-hand cut.

By comparing Eq.~\eqref{211116.3} with the  ERE
\begin{equation}
t(E)=\frac{1}{-1/a+1/2 r k^2 - i k},
\end{equation}
one obtains
\begin{equation}
%1/a=\frac{\lambda}{\MCDD-\Mth},\qquad r= -\frac{\lambda}{\mu(\MCDD-\Mth)^2}~.
1/a=\frac{\lambda}{\MCDD},\qquad r= -\frac{\lambda}{\mu \MCDD^2}~.
\end{equation}
If $\MCDD$ is far from threshold,
 $r$ would have a natural size around 1 fm, and the two meson constituents dominate;
otherwise when $\MCDD$ is vanishing, so that the CDD pole lies very close to threshold,
the compositeness will be a small number, which implies the non-molecule nature.

For the definition of compositeness ($X$) of a resonance, we adopt the one developed in
Ref.~\cite{Guo:2015daa}, adapted to the non-relativistic near-threshold case in Ref.~\cite{Zbpaper}.
Denoting by $g_k^2$ the residue in the variable $k$
of the partial-wave amplitude for a pole, either a bound state or resonance,
one has that $X=-ig_k^2$ for the former, while
$X=|g_k^2|$  for a resonance if $M_R>0$ (with $M_R$ the mass of the resonance measured from the threshold).

%which is applicable under the condition that $\sqrt{Re(s_R)}$ is larger than the lightest threshold,
%\begin{equation} \label{Eq:X}
%X=\left|\gamma^2\frac{dG_R}{ds_R}\right|,
%\end{equation}
%with $\gamma$ being the residue in the 2nd Riemann sheet for a resonance, and $G$ the two-point Green function, and $s_R=E_R^2$ denotes the pole position in $s-$plane.

We are now in position to obtain the theoretical formula for the event distributions, which are also the experimentally measured observables.
We define the signal function as
\begin{equation}
\label{231116.1}
\frac{d\hat M}{dE}=\frac{\Gamma_X|d(E)|^2}{2\pi|\alpha|^2}~,
\end{equation}
with $\alpha$ introduced such that in the vicinity of the pole position $d(E)\simeq \alpha/(E-E_p)$.
As done in Ref.~\cite{Braaten}, the background term, in the energy region considered, is
treated as a constant or a linear function.
The event distribution $N(E)$ is then expressed by the incoherent sum of the signal part and the background.
In all the cases, the overall normalization constants, the parameters characterizing the
background and the parameters $\lambda,\beta$ and $\MCDD$, which determine the signal function,
are free parameters, fitted to data.
In practice, however, there are too many free parameters, which prevents us to extract any definite conclusion by
reproducing the experimental data.
To reduce the numbers of free parameters, we discuss several interesting scenarios that drive to
different nature and pole structures for the $X(3872)$ (which may also represent feasible situations for other
near-threshold states).
As a result the number of free parameters in $d(E)$ for each case is 1 or 2, so that
one can obtain  for every scenario stable fits to data. The cases considered are:

\begin{itemize}
\item[Case 1:] Using the ERE with only the scattering length,
which predicts the $X(3872)$ as a pure composite bound state \cite{Braaten}.
%we make a combined fit to all the existing data, shown in Fig.~\ref{fig:resultI}.
%versus the separate fit in Ref.~\cite{Braaten}.

\item[Case 2:] Imposing that $t(E)$ (for $\Gamma_*=0$) has a double-virtual pole
  state  as the limiting zero-width case of a resonance,
one can express $\lambda$ and $\beta$ from $E_R$ and $\MCDD$.
  That is, there are two free parameters, denoted by Case 2.I (generating two virtual-state poles).
  Furthermore, if one imposes $\MCDD=-3E_R$ then a triple virtual state pole happens for $\Gamma_*=0$,
  and this is denoted as Case 2.II.

\item[Case 3:] Taking into account the coupled channel effects and imposing independence of the pole position
  with respect to the threshold of the charged or neutral $D\bar{D}^*$ pairs, so that the self-energy contribution due
  to re-scattering is necessarily suppressed.
  Then, the compositeness is very small at the level of only a few percent,
  which implies the ``elementary'' (or ``bare'') nature of this particle.
  Here, only $E_R$ determines the shape of $t(E)$, but
  we encounter a quadratic equation with two solutions which give rise to  Case 3.I  and Case 3.II.

  \end{itemize}

\begin{figure}[htbp]
\begin{center}
\vglue 0.5cm
\includegraphics[height=85mm,clip]{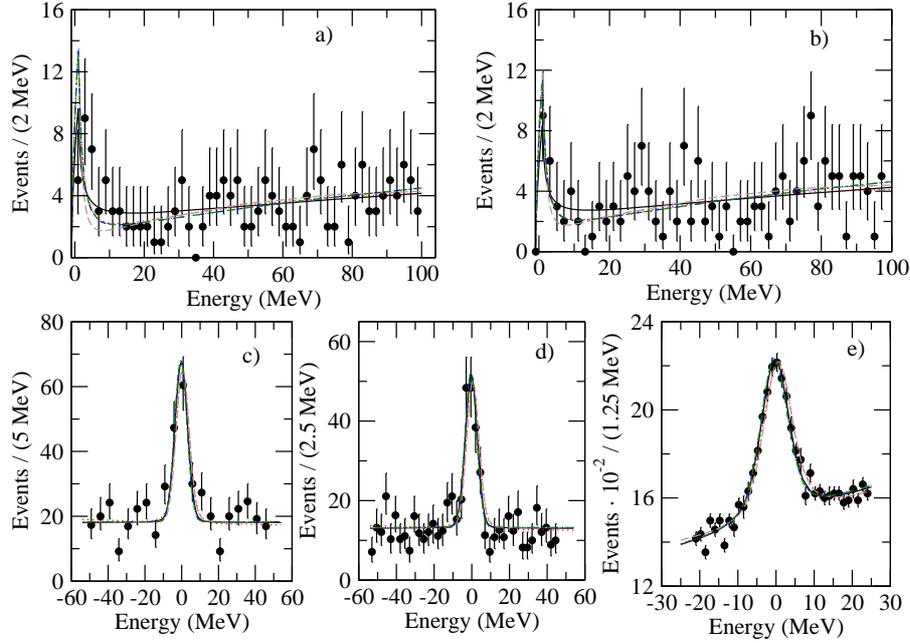}
\caption{Fit results  for the different
event distributions as a function of energy. Panels a) and b) denote
the mode $D^0\bar D^{*0}$ from BaBar \cite{BaBarD} and Belle
\cite{BelleD}, respectively. Panels c), d) and e) denote the mode
$J/\psi\pi^+\pi^-$ from BaBar \cite{BaBarJ}, Belle \cite{BelleJ},
and CDF \cite{CDF.211116.5} collaborations, respectively. The black
solid lines correspond to case 1.
 The CDD pole contribution is included in the expression of $d(E)$, and the rest of lines refer to cases making use of this
parameterization. Case 2.I is shown by the red dotted lines, and
case 2.II by the brown dashed lines, while cases 3.I and 3.II are given by
the blue dash-double-dotted and green dash-dotted lines, in
order. Part of the lines overlap so much that it is difficult to
distinguish between them in the scale of the plots.}
 \label{fig:resultI}
\end{center}
\end{figure}

The results that follow by fitting simultaneously  all the combined data
are shown in Fig.~\ref{fig:resultI}.
There one can see that all these curves describe data almost equally well.
The resulting pole locations are given in Table~\ref{tab:fitsummary}, indicating the compositeness $X$ for
those cases in which we can calculate it, as discussed above.

\section{Conclusion and discussion}
We propose to use a more general parameterization of a two-body scattering amplitude around threshold.
With this advanced tool, we derive the formula for the event distributions
 with the free parameters obtained by fitting to data, which is well reproduced.
The pole structures are analyzed, and it turns out that the $X(3872)$ can be interpreted as either bound or virtual state,
or even as an ``elementary''  state.
That is, the current data based line shapes is not enough for determining the properties of the $X(3872)$.
Let us notice especially the Case 2, which implies that the $X(3872)$
stems from a double- or triple-virtual state pole.
More higher-precision data or other observables are highly desired to further discriminate between those scenarios.

\begin{table}[htbp]{\footnotesize
 \tbl{Fits for all the cases including the pole positions and compositeness
($X$).}
{\begin{tabular*}{0.8\linewidth}{@{\extracolsep{\fill}}ccc}
\hline\hline
 Case   & Pole position [MeV]    & $X$   \\
\hline
1       &$-0.19_{-0.01}^{+0.01} - i \, 0.0325$   & $1.0$ \\
\hline
2.I     &$-0.36_{-0.10}^{+0.08} - i\, 0.18_{-0.02}^{+0.01}$   &  \\
        &$-0.70^{+0.11}_{-0.13} + i \, 0.17_{-0.01}^{+0.02}$  &  \\
\hline
2.II    &$-0.33_{-0.03}^{+0.04} - i\, 0.31_{-0.01}^{+0.02}$  &  \\
        &$-0.84^{+0.07}_{-0.05} + i \, 0.77_{-0.04}^{+0.03}$  & \\
        &$-1.67_{-0.08}^{+0.10} - i \,0.49_{-0.02}^{+0.02}$   &  \\
\hline
3.I     &$-0.50^{+0.04}_{-0.03}$ &$0.061_{-0.002}^{+0.003}$  \\
        &$-0.68^{+0.05}_{-0.03}$ &                            \\
\hline
3.II    &$-0.51^{+0.03}_{-0.01}$ &$0.158_{-0.001}^{+0.001}$  \\
        &$-1.06^{+0.05}_{-0.02}$ &                            \\
\hline\hline
\end{tabular*}}
 \label{tab:fitsummary}}
\end{table}

%\bibliographystyle{ws-procs9x6} % for numbered citation & references
%\bibliography{ws-pro-sample}

%Non BiBTeX users can list down their references as:

\end{document}